\documentstyle[preprint,pre,aps,amssymb,epsfig]{revtex}
\tightenlines

\begin{document}

\draft
\title{Phase diagram for morphological transitions of wetting films on
  chemically structured substrates}
\author{C. Bauer and S. Dietrich}
\address{Fachbereich Physik, Bergische Universit\"at Wuppertal,\\
D-42097 Wuppertal, Germany}

\maketitle

\begin{abstract}

Using an interface displacement model we calculate the shapes of thin
liquidlike films adsorbed on flat substrates containing a chemical stripe.
We determine the entire phase diagram of morphological phase
transitions in these films as function of temperature,
undersaturation, and stripe width.

\end{abstract}

\section{Introduction}

At present various experimental techniques allow one to provide
substrates with well-defined, microscopically small geometrical or chemical
structures or combinations thereof (see, e.g., Ref.~\cite{microstructuring}). These
structured substrates can serve as devices to guide and process tiny
amounts of liquids. Such ``microfluidics'' systems~\cite{microfluidics} are
promising for future applications, 
e.g., in chemistry and biology~\cite{chembiol1,chembiol2}. Although
the performance of these systems ultimately depends on their dynamic
properties, a thorough understanding of the corresponding structural
properties in thermal equilibrium is an important
prerequisite~\cite{sdsicily}. In this spirit as a
paradigmatic case we determine
the equilibrium structures of liquidlike wetting films forming at the
interface between a bulk vapor phase and a flat solid substrate
containing a single chemical stripe (Fig.~\ref{f:system}). Our aim is
to explore the entire phase diagram for the emerging lateral fluid
structures as function of temperature, undersaturation or pressure,
and stripe width.

In Ref.~\cite{bauerdietrichparry} it has been shown that a thin liquidlike film
adsorbed on such a substrate (Fig.~\ref{f:system}) can undergo an interesting
morphological phase transition (Figs.~\ref{f:tdm} and \ref{f:surface})
of the shape $l(x)$ of the liquid-vapor interface, depending on the
structures of the effective interface potentials $\Lambda_{\pm}(l)$
characterizing the corresponding \emph{homogeneous} substrates ``$+$''
and ``$-$'' (see, e.g., Fig.~\ref{f:pointa}). 
In accordance with Ref.~\cite{bauerdietrichparry} we
study a substrate designed such that the effective interface
potential $\Lambda_+(l)$ of the stripe part exhibits two
competing local minima (``I'' near and ``III'' further away
from the substrate surface) whereas the asymptotic 
film thickness $l_-$ far from the stripe is given by the single global
minimum (``II'') of the interface potential $\Lambda_-(l)$ of the embedding 
substrate (see Fig.~\ref{f:pointa}). In 
Ref.~\cite{bauerdietrichparry} the phase diagram for this
morphological phase transition has been determined, at constant
temperature $T$, as function of the stripe width $a$ and of the
undersaturation $\Delta\mu = \mu_0-\mu\geq0$ where $\mu_0(T)$ is the
chemical potential at liquid-vapor coexistence. The purpose of the
present study is to determine how this phase diagram for the
morphological phase transition within the $a$-$\Delta\mu$ plane
evolves as function of temperature. This additional information is
important in order to be able to put the morphological phase
transitions into the context of $T$-$\Delta\mu$ wetting phase
diagrams, including the prewetting line, on homogeneous
substrates~\cite{sdreview}. 

We emphasize that experimental techniques such as reflection
interference contrast microscopy (RICM)~\cite{wiegandetal} and atomic
force microscopy (AFM) in tapping mode~\cite{herminghaus} are capable
of imaging the liquidlike structures under consideration down to the
nm scale. Moreover, due 
to their relatively small spatial extensions the structures discussed
here are well accessible by Monte Carlo simulations as another means
for testing the theoretical predictions. For these envisaged
studies the knowledge of the full topology of the phase diagram as
function of $T$, $\Delta\mu$, and $a$ is very important.

\section{Theoretical model}

We use the following simple interface displacement model within which
the equilibrium interface profile $\bar{l}(x)$ minimizes the functional 
\begin{equation}\label{e:func}
\Omega_{\mathcal S}[l(x)] = \int_A dx\,dy\,\Lambda(x,l(x)) +
\sigma_{lg}\int_A dx\,dy\,\left(\frac{dl(x)}{dx}\right)^2
\end{equation}
where $A=L_xL_y$ is the area of the flat substrate surface located at
$z=0$. The first part takes into account the effective interaction
between the liquid-vapor interface and the substrate via the laterally
varying effective interface potential
$\Lambda(x,l) = \Delta\Omega_b\,l+\omega(x,l)$. The second
part in Eq.~(\ref{e:func}) is the leading-order term of a gradient
expansion of the surface free 
energy associated with the deviation of the liquid-vapor interface
profile from its flat configuration; $\sigma_{lg}$ is the
surface tension. $\Delta\Omega_b = \Delta\mu(\rho_l-\rho_g)+{\mathcal
  O}((\Delta\mu)^2)$ is the difference of the bulk free 
energy densities of the liquid and the vapor phase with number
densities $\rho_l$ and $\rho_g$, respectively. Alternatively, in terms of
the bulk pressure $p$, $\Delta\Omega_b$ is given approximately by the
Gibbs-Thomson expression $k_BT\rho_l\,\ln(p_{sat}/p)$ where $p_{sat}$
is the saturated vapor pressure. At liquid-vapor coexistence
$\Lambda(x,l)$ is given by $\omega(x,l)$ which is of the form
$\sum_{i\geq2}a_i(x)/l^i$. For reasons of
simplicity we assume that the lateral variation of $\omega(x,l)$ is
steplike: $\omega(x,l) = 
\Theta(|x|-a/2)\omega_-(l)+\Theta(a/2-|x|)\omega_+(l)$ with the effective
interface potential $\omega_+$ ($\omega_-$) of a \emph{homogeneous} substrate
composed of ``$+$'' (``$-$'') particles. 
In Refs.~\cite{bauerdietrich2} and \cite{bauerdietrich1} we have shown
that the functional in Eq.~(\ref{e:func}) -- despite its simplicity --
allows one to determine reliably the morphology of liquidlike films on structured
substrates, and we have presented a derivation of this
square-gradient functional from a microscopic density
functional theory. We also found (see Fig.~11 in
Ref.~\cite{bauerdietrich2}) that the aforementioned steplike variation
of $\omega(x,l)$ as opposed to its actual smooth lateral variation
represents a rather reliable approximation. The form of
$\omega(x,l)$ follows from considering as the interaction potential between
the fluid particles
a Lennard-Jones potential with a depth $-\epsilon$ and
an effective particle diameter $\sigma$. Moreover,
also the interaction between the fluid and the substrate particles
is modeled by a Lennard-Jones potential (compare Ref.~\cite{bauerdietrich2}).
Since the substrate is translationally invariant in the $y$ direction
the effective interface potential depends only on $x$, so that
$\bar{l}(x)$ is also translationally invariant in the $y$ direction.
The interaction potential parameters of the stripe part (``$+$'') are the same
as for the ``$+$'' substrate described in Fig.~11 in Ref.~\cite{bauerdietrich2}, and
for the embedding ``$-$'' substrate part we use the parameters for the ``$+$''
substrate part in Fig.~7 in Ref.~\cite{bauerdietrich2} multiplied by
$0.9102$. The temperature dependence of the effective interface
potentials $\Lambda_{\pm}(l)$ is determined by systematically keeping track of the
temperature dependence of the bulk liquid and vapor densities $\rho_l$
and $\rho_g$ which enter into $\Lambda_{\pm}(l)$ as discussed in Sec.~II in
Ref.~\cite{bauerdietrich2}. For the case studied here the temperature dependence of
$\omega_+(l)$ for the stripe part 
is such that a homogeneous ``$+$'' substrate exhibits a first-order
wetting transition at $T_w^* = k_BT_w/\epsilon \approx 
1.102$ and prewetting transitions along the prewetting line
$\Delta\mu_{pre}(T)$, $T>T_w$ and $\Delta\mu>0$, that extends into the vapor
phase region of the phase diagram (thick dashed line denoted as ``P''
in Fig.~\ref{f:tdm}). $\omega_-(l)$ for the embedding substrate is also
temperature dependent such that the asymptotic film thickness $l_-$ grows
with increasing $T$ and shrinks with increasing
$\Delta\mu$. As it turns out, however, the
exact temperature dependence and the magnitude of $l_-$ are not important for
the qualitative features of the phase diagram for the morphological transitions. 

The systems under consideration here differ from those studied in
Ref.~\cite{gauetal} in two important aspects. First, here the
liquidlike films are so thin that they are completely under the
influence of the substrate potential which determines, inter alia, the
effective interface potential. Secondly, we consider a grand
canonical ensemble without a volume constraint for the liquid phase so
that the liquidlike films are not subject to instabilities along the
$y$ direction as observed in, e.g., Ref.~\cite{gauetal}.

Using a numerical relaxation technique~\cite{numrec} we solve the two-point boundary
value problem for the Euler-Lagrange equation 
\begin{equation}\label{e:ele}
\sigma_{lg}\frac{d^2 l(x)}{dx^2} = \Delta\Omega_b +
\left.\frac{\partial\omega(x,l)}{\partial l}\right|_{l=l(x)}
\quad\mbox{with}\quad l(x\to\pm\infty) = l_-
\end{equation}
which follows from Eq.~(\ref{e:func}) by functional differentiation
with respect to $l(x)$. Equation~(\ref{e:ele}) exhibits the structure of a
one-dimensional classical mechanical equation of motion in a
time-dependent external potential. With the stripe width $a$ and the
temperature $T$ fixed, for a wide range of undersaturations
there are two solutions of Eq.~(\ref{e:ele}) which correspond to local
minima of the functional $\Omega_{\mathcal S}$. One of the solutions
is closely bound to the stripe whereas the other solution is further away
or even repelled from the stripe as described in
Ref.~\cite{bauerdietrichparry}. Figures~\ref{f:pointa}--\ref{f:pointd}
show pertinent examples for the effective potentials $\Lambda_{\pm}$ of the
two substrate materials and the two appertaining solutions of
Eq.~(\ref{e:ele}) at three different points of the phase diagram:
 Figs.~\ref{f:pointa}, \ref{f:pointb}, \ref{f:pointc}, and \ref{f:pointd}
correspond to the points $(T,\Delta\mu)$ indicated by ``a'', ``b'',
``c'', and ``d'', respectively, in the phase diagram shown in
Figs.~\ref{f:tdm} and \ref{f:surface}. The equilibrium 
solution $\bar{l}(x)$ for the profile corresponds to the global minimum of
$\Omega_{\mathcal S}$, and the other solution is metastable. At a certain
value $\Delta\mu_t$, for which $\Omega_{\mathcal S}$ has the same value for both
solutions, a phase transition from one interfacial configuration to
the other takes place. This transition is first order because the
derivatives of $\Omega_{\mathcal S}(\bar{l}(x))$ with respect to
$\Delta\mu$ are discontinuous. 

As already shown in Ref.~\cite{bauerdietrichparry}, in Eq.~(\ref{e:func})
the two contributions to $\Omega_{\mathcal S}$ compete with each other
in minimizing the whole functional.
Depending on the special choices for $a$, $T$, and $\Delta\mu$ the cost
in free energy for increasing the liquid-vapor interface area
is overcompensated by the gain in free energy which follows from
occupying the deeper minimum (I) of $\Lambda_+$ near the wall, so
that the bound configuration has the lower free energy. In the
opposite case this gain
in free energy is too small to compensate the cost in free
energy from the increased area of the liquid-vapor interface, leading
to a less bound or even repelled interface profile.

\section{Discussion of the phase diagram}

In Fig.~\ref{f:tdm} the lines $\Delta\mu_t(T;a) = \mu_0(T)-\mu_t(T;a)$
of phase coexistence between the bound and 
the repelled configuration for different stripe widths $a$ are
presented. These lines are vertical cross-sections of the full phase
diagram shown in Fig.~\ref{f:surface}. For a given stripe width $a$
the bound (repelled) solution is stable 
for $-\Delta\mu<-\Delta\mu_t(T;a)$ ($-\Delta\mu>-\Delta\mu_t(T;a)$),
i.e., below (above) 
the corresponding line of phase coexistence. The triangles indicate
the intersections $T_0(a)$ between the coexistence lines
$\Delta\mu_t(T;a)$ and the bulk liquid-vapor coexistence line
$\Delta\mu=0$. At $T_0(a)$ the morphological phase transition occurs
at liquid-vapor coexistence, forming the line denoted as $\alpha$ in
Fig.~\ref{f:surface}. Upon increasing the stripe 
width $a$ the lines of coexistence are shifted towards the prewetting
line ``P'' without touching or crossing it for any finite value of
$a$. (This means that for large $a$ the lines of coexistence do not
end and reappear at the prewetting line such as, e.g., the
prefilling transition lines in a wedge-shaped groove for large
opening angles of the wedge~\cite{rejmer}.) For constant
undersaturation $\Delta\mu$ we find
$T_{pre}(\Delta\mu)-T_t(\Delta\mu;a\to\infty)\sim a^{-1}$ for the
difference between the transition temperatures $T_t(\Delta\mu;a)$ for
the morphological and the prewetting transition. In the limit $a\to\infty$
the morphological phase transition tallies with the prewetting
transition on the homogeneous ``$+$'' substrate. 

On the mean-field level considered hitherto each coexistence curve
$\Delta\mu_t(T;a)$ is a line of first-order transitions ending in a
critical point $(T_c(a),\Delta\mu_c(a))$ (denoted as full diamonds in
Fig.~\ref{f:tdm} and forming the line denoted as $\gamma$ in
Fig.~\ref{f:surface}) such that at $T>T_c(a)$ or
$-\Delta\mu<-\Delta\mu_c(a)$ for the given stripe width $a$ there is no
morphological transition but a smooth variation from the bound to the
repelled solution. As demonstrated by Fig.~\ref{f:tdm} and its inset
the positions of the critical points 
exhibit a nontrivial and nonmonotonous dependence on the stripe width
$a$; there is no simple criterion for the effective interface potential
$\Lambda_{\pm}$ which allows one to predict the corresponding line of
critical points (see the line $\gamma$ in Fig.~\ref{f:surface}) as function of $a$. 
As already pointed out in detail in Ref.~\cite{bauerdietrichparry}
interface fluctuations along the $y$ direction in this effectively
one-dimensional stripe configuration actually smear out
the sharp first-order morphological phase transitions and thus eliminate 
the critical points~\cite{finitesize}. In Fig.~\ref{f:tdm} for
$a=5\sigma$ the width of this fluctuation-induced smooth transition region
is indicated by the thin dashed-dotted lines. For
$a\geq10\sigma$ these fluctuation effects are already negligibly
small, apart from the close vicinity of the critical points which are
still eliminated for any finite $a$. However,
in the limit $a\to\infty$ the coexistence lines $\Delta\mu_t(T;a)$
merge with the prewetting line ``P'' (see Fig.~\ref{f:surface}) associated
with the homogeneous ``$+$'' substrate, which does have a genuine
critical point ``C'' beyond mean field theory.

The three thermodynamic states denoted as ``a'', ``b'', and ``c'' in
Figs.~\ref{f:tdm} and \ref{f:surface} lead to the interface configurations shown in 
Figs.~\ref{f:pointa}, \ref{f:pointb}, and
\ref{f:pointc}, respectively. The corresponding effective interface
potentials $\Lambda_{\pm}(l)$ differ with respect to the number and the relative
positions of their local minima. For temperatures
$T<T_{pre}$ and small undersaturations $\Delta\mu$, $\Lambda_+(l)$ exhibits
two local minima (I and III); an example for this case is shown in 
Fig.~\ref{f:pointa} appertaining to the thermodynamic state denoted as
``a'' in Fig.~\ref{f:tdm}. This case is analogous to that described in 
Ref.~\cite{bauerdietrichparry}. At the thin dotted line
$\Delta\mu_i(T)$ denoted as ``i'' in
Fig.~\ref{f:tdm} the minimum III of $\Lambda_+$ far from the wall and the 
maximum of $\Lambda_+$ merge, forming a saddle point; $\Delta\mu_i(T)$
ends at the prewetting critical point ``C''. For 
$-\Delta\mu\leq-\Delta\mu_i(T)$ there is only one local minimum (I) of $\Lambda_+$.
An example for this latter situation is shown in Fig.~\ref{f:pointb}
corresponding to the thermodynamic state denoted as ``b'' in
Fig.~\ref{f:tdm}. Surprisingly the coexistence lines $\Delta\mu_t(T;a)$
for the morphological phase transition 
extend below this line ``i''. This shows that the presence of an energy 
barrier and of a second local minimum of $\Lambda_+$ is not a necessary condition
for the occurrence of the morphological phase transition described here and
in Ref.~\cite{bauerdietrichparry}. The thin dotted line
$\Delta\mu_{iii}(T)$ denoted as ``iii'' is the
line at which, analogously, the minimum I near the wall and the maximum of
$\Lambda_+$ merge, forming a saddle point. Both lines ``i'' and
``iii'' meet and end at the critical point ``C'' of the prewetting transition of a 
homogeneous ``$+$'' substrate where both minima I and III and the maximum
of $\Lambda_+$ merge, leaving a single minimum.

At the thin short-dashed line $\Delta\mu_{ii}(T)$ denoted as ``ii'' in
Fig.~\ref{f:tdm} the positions of the 
second minimum III of $\Lambda_+$ and of the minimum II of $\Lambda_-$ coincide.
Along this line one of the solutions of Eq.~(\ref{e:ele}) is completely
flat: due to $d\Lambda_{\pm}/dl\,(l=l_-)=0$ Eq.~(\ref{e:ele}) leads to
the trivial solution $l(x)\equiv l_-$ with $d^2l(x)/dx^2\equiv0$. An example 
for this case is shown in Fig.~\ref{f:pointc} which corresponds to the
thermodynamic state denoted as ``c'' in Fig.~\ref{f:tdm}. The line ``ii''
separates the region where 
the repelled solution is bent towards the stripe
($-\Delta\mu<-\Delta\mu_{ii}(T)$) from that where it is bent away from the
stripe ($-\Delta\mu>-\Delta\mu_{ii}(T)$). We note that the fact that one
of the solutions is completely flat is an artifact of our simplifying
model assumption that $\omega(x,l)$ varies steplike as function of
$x$. In a more realistic model with a smooth lateral variation of the effective 
interface potential (as used in Ref.~\cite{bauerdietrichparry}) the corresponding
solution would be almost flat with a small curvature. In the system considered
here the asymptotic film thickness $l_-$, i.e., the position of minimum II, is
always larger than the position of minimum I. However, for low temperatures
there is a line at which the minimum II of $\Lambda_-$ and the
\emph{maximum} of $\Lambda_+$ coincide. This occurs along the thin long-dashed line
$\Delta\mu_{iv}(T)$ denoted as ``iv'' in Fig.~\ref{f:tdm}. This line is the
continuation of line ``ii'' for lower temperatures $T$ starting at the point
where the line ``ii'' meets the line ``i'' tangentially. Along this line, too,
the repelled solution of Eq.~(\ref{e:ele}) is constant, i.e.,
$l(x)\equiv l_-$. As the line ``ii'', also line ``iv'' separates two
regions with different curvature behavior of the repelled solution.

For temperatures $T>T_{pre}$, i.e., on the right side of the prewetting line P,
the minimum III of $\Lambda_+$ is deeper than the minimum I. The
difference between the asymptotic film thickness $l_-$ corresponding
to the minimum II and the deeper
minimum III is also smaller than that between $l_-$ and the local minimum I
near the wall. Therefore both contributions to $\Omega_{\mathcal S}$
are larger in the case that the interface follows minimum I as
compared to the case that it follows minimum III so that for
$T>T_{pre}$ the bound solution is 
always metastable. This situation is shown in Fig.~\ref{f:pointd} which
corresponds to the thermodynamic state denoted as ``d'' in
Fig.~\ref{f:tdm}. If $T\geq T_{iii}(\Delta\mu)$ 
$\Lambda_+$ exhibits only one local minimum (III) and therefore there is
only one solution of Eq.~(\ref{e:ele}), namely the repelled one. 
 
\section{Conclusion}

\begin{figure}
\begin{center}
\epsfig{file=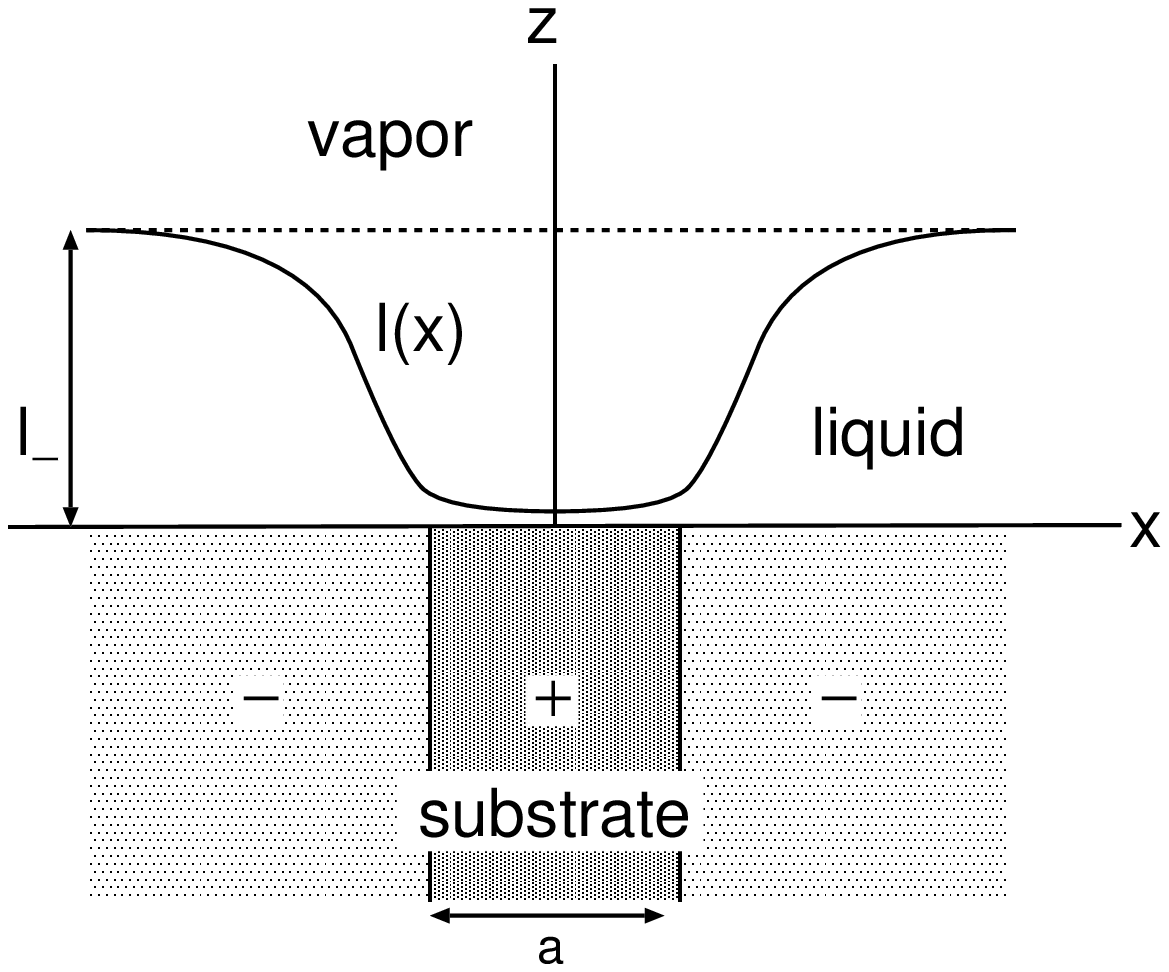, width = 10cm}
\end{center}
\caption{\label{f:system}
Schematic cross-section of the morphology of a liquidlike film covering a
planar substrate which contains a slab of different material. In top
view the substrate forms a stripe composed of ``$+$'' particles 
extending from $x=-a/2$ to $x=a/2$. The
substrate is translationally invariant in the $y$ direction, with its
surface located at $z=0$. $l_-=l(|x|\to\infty)$ is the equilibrium film thickness
corresponding to a homogeneous substrate composed of ``$-$''
particles. The stripe (embedding) part exerts an effective interface potential
$\Lambda_+$ ($\Lambda_-$) on the liquid-vapor interface at the
vertical distance $z=l(x)$. In the case shown here the ``$-$''
substrate prefers a thicker wetting film than the ``$+$'' substrate.}
\end{figure}

\begin{figure}
\begin{center}
\epsfig{file=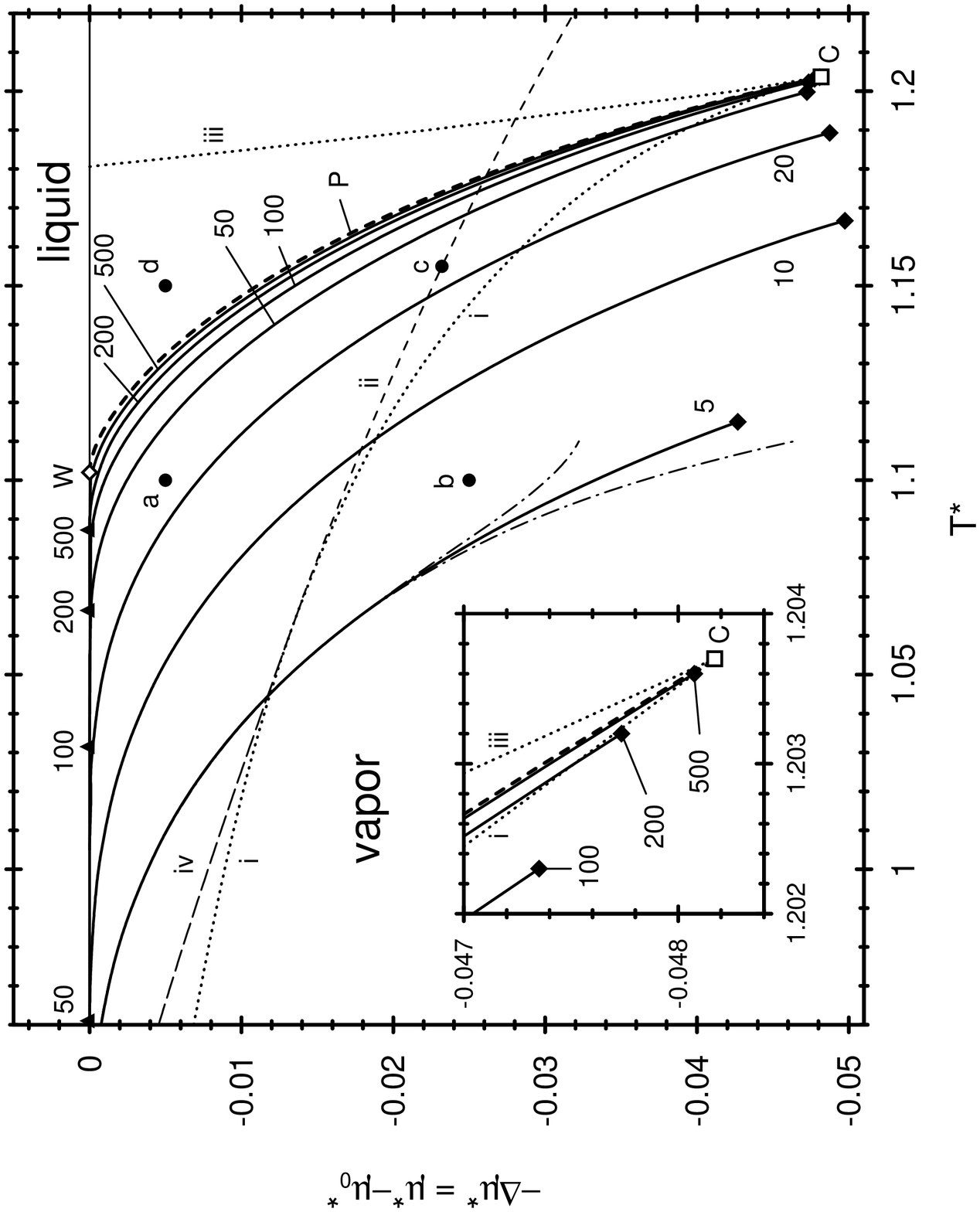, width=16cm, bbllx=20, bblly=60,
  bburx=580, bbury=770} 
\end{center}
\caption{\label{f:tdm}
$T$-$\Delta\mu$ phase diagram for the morphological transitions of a
liquidlike layer on a chemical stripe (compare Fig.~\ref{f:system});
$T^* = k_BT/\epsilon$ and $\mu^* = \mu/\epsilon$. $\Delta\mu=0$ is the
bulk liquid-vapor coexistence line. For the present model the bulk
critical temperature is $T_c^*\approx1.412$; the estimated triple
point temperature is $T_t^*\approx0.8$. The \emph{homogeneous}
substrate composed of ``$+$'' particles, i.e., corresponding to the
stripe, supports a first-order wetting transition at coexistence
denoted by ``W'' ($\lozenge$). The corresponding prewetting line ``P''
(thick dashed line) is attached tangentially and ends at a prewetting
critical point ``C'' ($\square$). The effective interface
potentials $\Lambda_{\pm}$ and, as examples, the equilibrium interface
profiles at the thermodynamic states $\bullet$ denoted as ``a'', ``b'', ``c'', and
``d'' are shown in Figs.~\ref{f:pointa}--\ref{f:pointd} for $a/\sigma
\approx 27.5$, $6.4$, $26$, and $a/\sigma=50$, respectively. For ``a'',
``b'', and ``c'' these values for the width $a$ are those for which in these
thermodynamic states the morphological phase transition takes
place. The coexistence lines $\Delta\mu_t(T;a)$ (full lines) of the
bound and the repelled solutions for fixed stripe width $a$ are
accompanied by the appertaining value $a/\sigma$. For
$-\Delta\mu>-\Delta\mu_t$ the repelled solution and for 
$-\Delta\mu<-\Delta\mu_t$ the bound solution is the stable one. The
lines of coexistence intersect the axis $\Delta\mu=0$ at the points $T_0(a)$
marked as $\blacktriangle$. $\Delta\mu_t(T;a)$ joins the bulk
liquid-vapor coexistence curve tangentially with $\Delta\mu_t(T\to
T_0(a);a)\sim (T-T_0(a))^{\delta}$; the numerical data indicate that
$\delta$ is larger than the corresponding exponent $\delta=3/2$ for the prewetting
line. Due to this tangential behavior the intersections
$T_0(a)$ occur at lower temperatures than expected visually on the
present scale. Therefore we have separately labeled the intersections $T_0(a)$ by
the corresponding values for $a/\sigma$. All lines of coexistence end
at a critical point denoted by $\blacklozenge$. The dashed-dotted
lines accompanying the line for $a=5\sigma$ show the region within
which the fluctuation-induced rounding of the first-order phase
transition takes place; for $a\geq10\sigma$ this rounding effect is
not visible on the present scale. The inset magnifies the region
around the prewetting critical point ``C''. The meaning of the lines
``i'', ``ii'', ``iii'', and ``iv'' is discussed in the main text.}
\end{figure}

\begin{figure}
\begin{center}
\epsfig{file=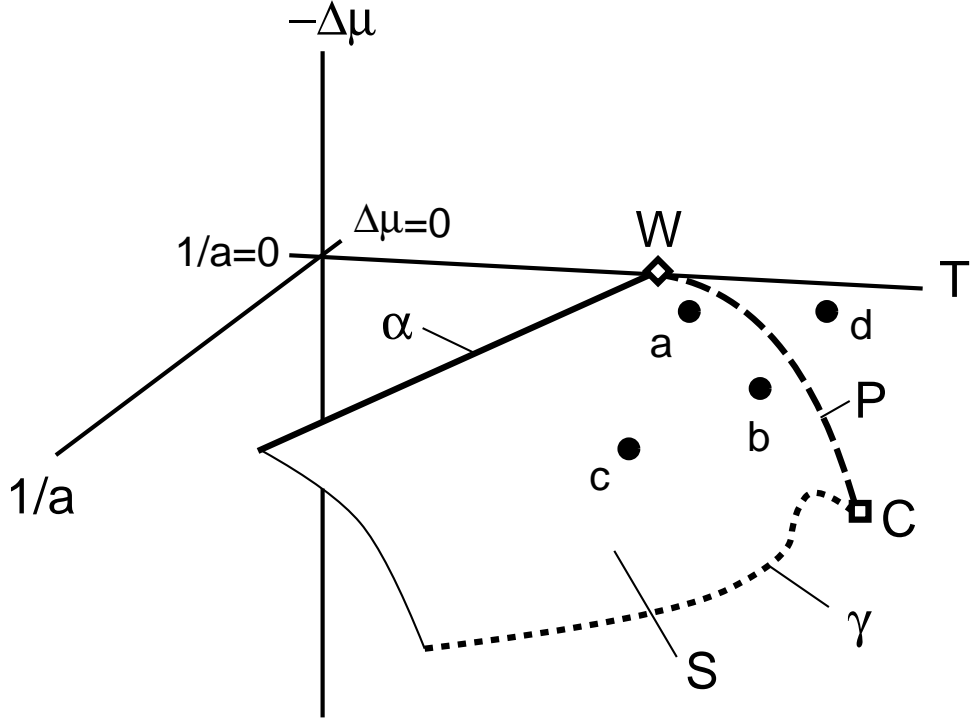, width = 13cm}
\end{center}
\caption{\label{f:surface}
The thermodynamic states at which the morphological phase transitions
take place are located on the surface ``S''. The lines of coexistence
shown as full lines in 
Fig.~\ref{f:tdm} are slices through this surface ``S'' at constant values of
$1/a$. ``S'' is a surface of first-order transitions and separates the
region where the repelled solution is stable 
(above ``S'') from the region where the bound solution is stable (below ``S''). It
is bounded by the prewetting line (dashed line ``P'') which lies in
the plane $1/a=0$, by the line $\gamma$ (dotted line) of critical points of the
morphological phase transitions, and by the line $\alpha$ which lies
in the plane $\Delta\mu=0$ indicating the
loci $(T_0(a),1/a)$ of the morphological phase transitions at
bulk liquid-vapor coexistence. ``W'' and ``C'' indicate the
first-order wetting transition on a homogeneous ``$+$'' substrate
($\lozenge$) and the prewetting critical point ($\square$),
respectively; both lie in the plane $1/a=0$. Indicated by the thin
line in the front, ``S'' extends out to larger values 
of $1/a$ where, however, it is increasingly smeared out by the
fluctuation-induced rounding of the morphological phase transition
(compare Fig.~\ref{f:tdm}). Due to
$T_{pre}(\Delta\mu)-T_t(\Delta\mu;a\to\infty)\sim
a^{-1}$, in terms of $1/a$ ``S'' is approximately a ruled surface: one of the two
principal curvatures is very small, vanishing exactly in the limit
$1/a\to0$. The thermodynamic states denoted by ``a'', ``b'' and ``c''
($\bullet$, compare Fig.~\ref{f:tdm}) lie on the surface ``S'' whereas the
thermodynamic state ``d'' is located above ``S''.}
\end{figure}

\begin{figure}
\begin{center}
\epsfig{file=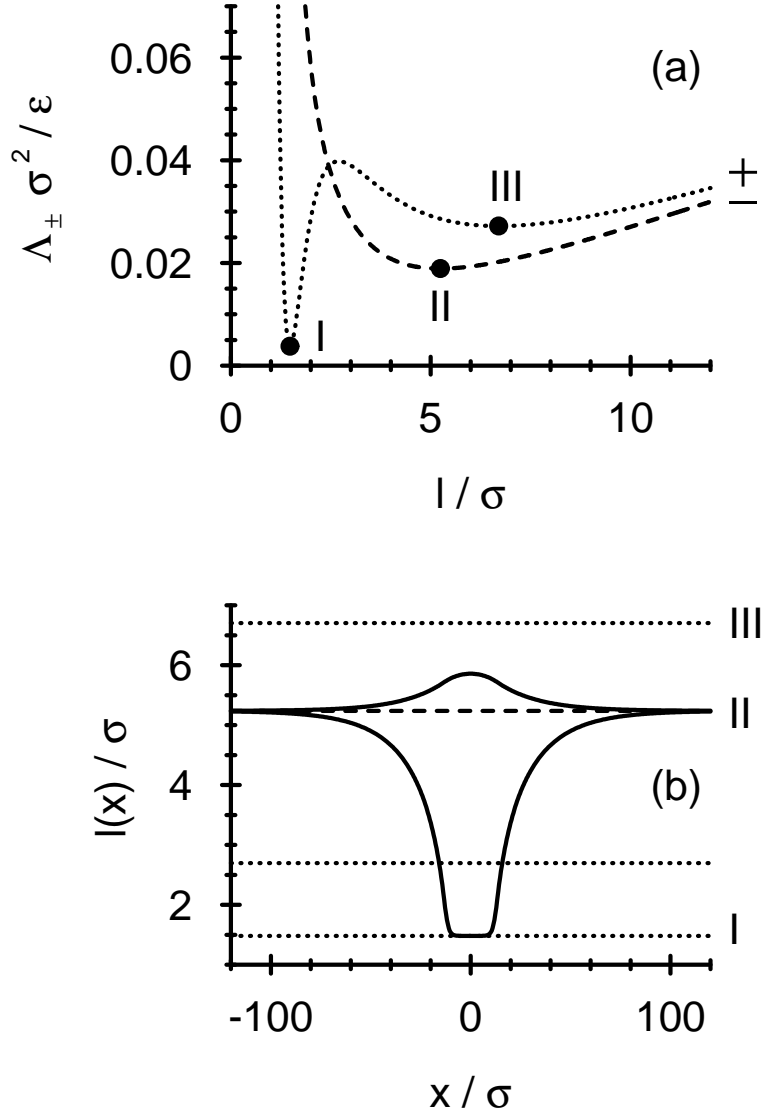, width = 10cm, bbllx = 55, bblly = 205,
  bburx = 410, bbury = 770}
\end{center}
\caption{\label{f:pointa}
(a) Effective interface potential $\Lambda_+$ of the stripe part
(dotted line) and $\Lambda_-$ of the embedding substrate (dashed
line) for the thermodynamic state denoted as ``a'' in
Figs.~\ref{f:tdm} and \ref{f:surface} ($T^* = 1.1,\Delta\mu^* =
0.005$). I, II, and III indicate the 
local minima of $\Lambda_{\pm}$. The global minimum
of $\Lambda_{\pm}$ corresponds to the equilibrium film thickness $l$ on a
homogeneous ``$\pm$'' substrate. $\Lambda_{\pm}(l\to\infty)$ increases
linearly with the slope
$(\rho_l-\rho_g)\Delta\mu$. (b) Equilibrium liquid-vapor interface profiles
for the same thermodynamic state as in (a) and for the stripe width
$a \approx 27.5\sigma$ for which at this value of $T$ and $\Delta\mu$ the
morphological phase transition takes place (see
Fig.~\ref{f:surface}). Therefore the value of 
the functional $\Omega_{\mathcal S}$ is the same for both
interface profiles shown here. The dotted (dashed) lines indicate the
positions of the extrema of $\Lambda_+$ ($\Lambda_-$).}
\end{figure}

\begin{figure}
\begin{center}
\epsfig{file=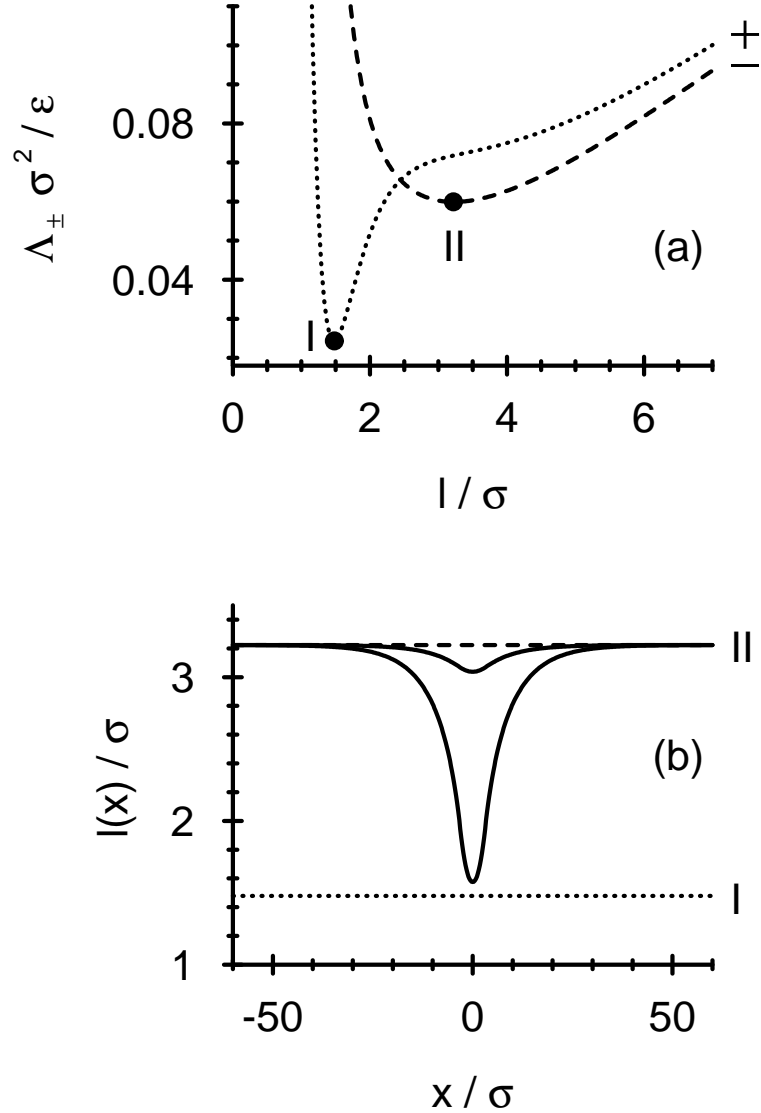, width = 10cm, bbllx = 55, bblly = 205,
  bburx = 410, bbury = 770}
\end{center}
\caption{\label{f:pointb}
(a) Same as in Fig.~\ref{f:pointa}(a), but for $T^* = 1.1$ and $\Delta\mu^* =
0.025$, i.e., for the thermodynamic state denoted as ``b'' in
Figs.~\ref{f:tdm} and \ref{f:surface}. The equilibrium interface
profiles shown in (b) correspond 
to $a \approx 6.4\sigma$, for which at the given thermodynamic state
the morphological transition takes place (see
Fig.~\ref{f:surface}). In contrast to the situation 
shown in Fig.~\ref{f:pointa} 
the repelled solution is bent towards the substrate and there is only
one local minimum (I) of $\Lambda_+$.}
\end{figure}

\begin{figure}
\begin{center}
\epsfig{file=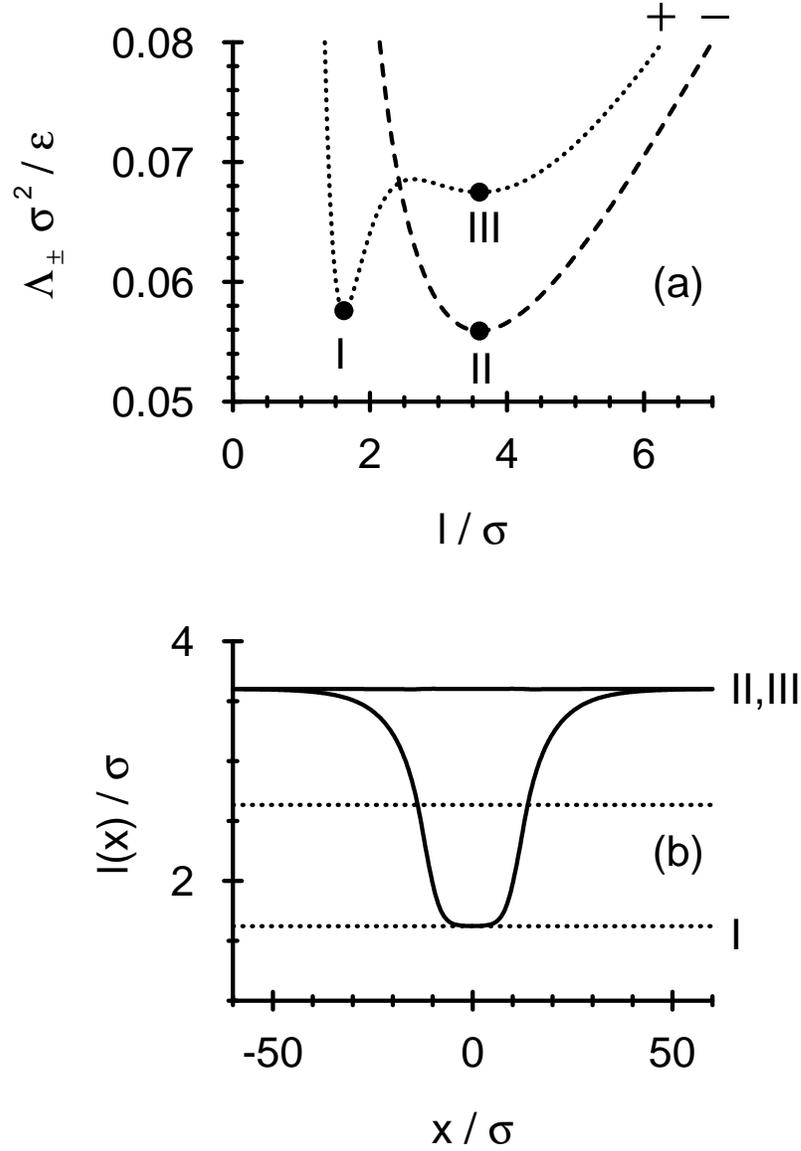, width = 10cm, bbllx = 55, bblly = 205,
  bburx = 410, bbury = 770}
\end{center}
\caption{\label{f:pointc}
(a) Same as in Fig.~\ref{f:pointa}(a), but for $T^* = 1.155$ and
$\Delta\mu^* \approx 0.0232$, i.e., for the thermodynamic state
denoted as ``c'' in Figs.~\ref{f:tdm} and \ref{f:surface} which is
located on the line ``ii''. The 
equilibrium interface profiles shown in (b) correspond to $a \approx
26\sigma$, for at which the given thermodynamic state 
the morphological transition takes place (see Fig.~\ref{f:surface}). The
positions of the minima II of $\Lambda_+$ and III of $\Lambda_-$
coincide so that the repelled interface shown in (b) is flat:
$l_{rep}(x) \equiv l_-$ where $l_-$ is the position of both minima II
and III.} 
\end{figure}

\begin{figure}
\begin{center}
\epsfig{file=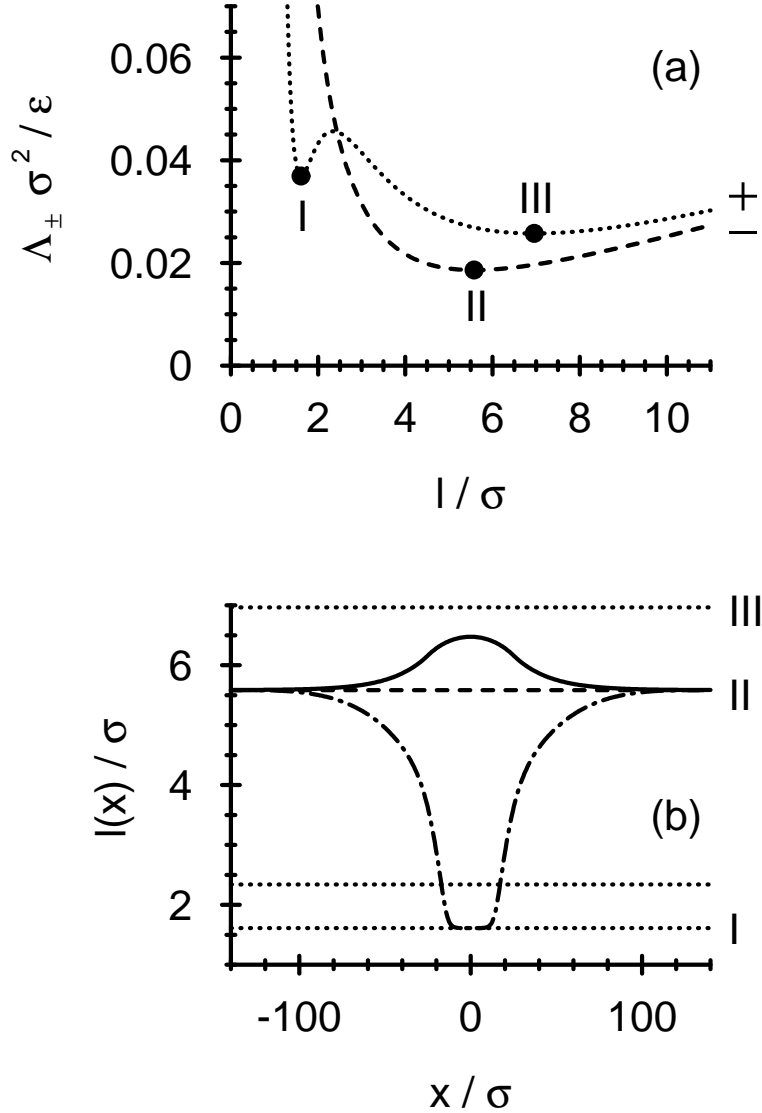, width = 10cm, bbllx = 55, bblly = 205,
  bburx = 410, bbury = 770}
\end{center}
\caption{\label{f:pointd}
(a) Same as in Fig.~\ref{f:pointa}(a), but for $T^* = 1.15$ and $\Delta\mu^* =
0.005$, i.e., for the thermodynamic state denoted as ``d'' in
Figs.~\ref{f:tdm} and \ref{f:surface}. (b)
displays the interface profiles for this state and for $a =
50\sigma$. In this case the
minimum III of $\Lambda_+$ is deeper than the local minimum I because of
$T>T_{pre}$. Therefore for all temperatures within the range
$T_{pre}(\Delta\mu)<T<T_{iii}(\Delta\mu)$, such as the one chosen here, the repelled
solution (full line) is the equilibrium profile (see
Fig.~\ref{f:surface}) whereas the bound solution
(dashed-dotted line) is metastable. The line $T_{iii}(\Delta\mu)$ is
shown as the dotted line ``iii'' in Fig.~\ref{f:tdm}.}
\end{figure}

Within an interface displacement model based on a microscopic density
functional theory we have calculated the entire phase
diagram (Figs.~\ref{f:tdm} and \ref{f:surface}) of morphological phase
transitions of wetting films on a substrate with a chemical 
stripe (Fig.~\ref{f:system}). This phase diagram and the corresponding
equilibrium interface profiles shown in
Figs.~\ref{f:pointa}--\ref{f:pointd} elucidate the 
dependence of the morphological phase transitions 
on the thermodynamic variables temperature and
undersaturation (or, equivalently, pressure) as well as on the stripe
width. The morphological phase transitions have been put into the
context of the prewetting transitions occuring on a homogeneous
substrate formed by particles of the stripe material.

\acknowledgements

We gratefully acknowledge financial support by the German
Science Foundation within the Special Research Initiative
\emph{Wetting and Structure Formation at Interfaces}.

\newpage

\end{document}